\documentclass[12pt]{article}
\usepackage{epsfig}
\usepackage{amsmath}
\usepackage{hhline}
\usepackage{amssymb}
\usepackage{times}

\newlength{\dinwidth}
\newlength{\dinmargin}
\newcommand{\GeV}{\rm GeV}

\newcommand{\F}{$ F_{2}(x,Q^2)\,$}

\begin{document}

\pagestyle{empty}
\begin{titlepage}

\vspace*{0.5cm}
\begin{verbatim}
DESY 02-036
March 2002                                   ISSN 0418-9833
\end{verbatim}
\vspace*{2cm}
\begin{center}
  \Large
  {\bf Self-similar Properties of the Proton Structure at low ${\mathbf x}$}
  \vspace*{1cm}

  {\Large T.~La\v{s}tovi\v{c}ka } \\
{\small Deutsches Elektronen-Synchrotron DESY, Zeuthen} \\
{\small Charles University, Faculty of Mathematics and Physics, Prague} \\

\end{center}
\vspace*{2cm}
\begin{abstract}

\noindent Self-similar properties of proton structure in the kinematic region 
of low Bjorken $x$ are introduced and studied numerically. A description of 
the proton structure function \mbox{\F} reflecting self-similarity is 
proposed with a few parameters which are fitted to recent HERA data. The 
specific parameterisation provides an excellent description of the data 
which cover a region of four momentum ~transfer squared, 
\mbox{$0.045 \le Q^2 \le 120~\GeV^2$}, and of
Bjorken $x$, 
\mbox{$6.2 \cdot 10^{-7} \le x \le 0.01$ }.

\end{abstract}


\end{titlepage}
\pagestyle{plain}


\section{Introduction}
\label{intro}
Recent measurements of the H1 \cite{thepaper} and ZEUS \cite{f2zeus} 
collaborations at HERA enable to study the proton structure in the 
region of low $Q^2 \lesssim 1~{\rm GeV}^2$ where perturbative QCD has 
to face computation difficulties arising from the increase of the 
strong coupling constant $\alpha_s(Q^2)$. 
Nevertheless, there is a number of approaches to describe the transition to 
low $Q^2$ at small $x$ together with the region, \mbox{$Q^2 > 1~{\rm GeV}^2$},
which is described by perturbative QCD very well. 
Such attempts involve Reggeon exchange ideas \cite{pomeron}, dipole 
interactions \cite{dipol}, vector meson dominance (VMD) \cite{VMD}, 
efficient parametrisations \cite{Haidt} and others.

This letter presents a different point of view, based on the idea that the 
proton structure at low $x$ is of fractal nature. Using the fractal dimension 
concept \cite{fractals}, a simple parametrisation of the proton structure 
function \F is obtained with a few well defined parameters. A numeric study 
is made using recent small $x$ HERA data, for $Q^2$ between $0.045~{\rm GeV}^2$
and $120~{\rm GeV}^2$.


\section{Fractal Dimension}
\label{dimension}
\begin{figure}[h]
\centering\epsfig{file=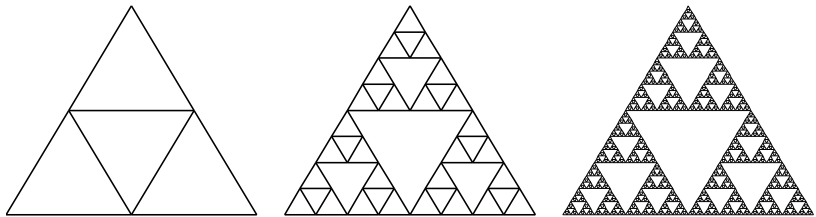,width=12.0cm,bbllx=30pt,bblly=75pt,
bburx=460pt,bbury=200pt}
\caption{Sierpinski gasket fractal in iterations No. 1, 3 and 6 
(from left). Iteration No. 1 corresponds to the {\it seed image} which is 
arbitrary while the iteration always converges to the same object.} 
\label{gasket}
\end{figure}

The concept of fractal dimension requires to understand what is meant by 
{\it dimension}. In non-fractional dimensions a number of dimensions 
corresponds to a number of independent directions in a corresponding 
coordinate system. For example, a line has obviously one dimension, a square 
two and a cube has three dimensions. The dimension of the Sierpinski 
gasket \cite{sierpinski}, shown in Fig. \ref{gasket}, needs a more general 
definition.

The cube, square and line are {\it self-similar} objects:
when a line is broken in the middle two lines are obtained, each of 
half length. By magnifying one of them by a factor of two the original line 
is rebuilt. The same may be done by dividing a square into four small squares 
or a cube into smaller cubes. For example, when the 
magnification factor for a square is 3 the number of smaller squares 
will be $3^2=9$, for a cube the same magnification gives $3^3=27$ cubes.
In general, when $M$ is the magnification factor, then the number of 
objects will be $M^{\cal D}$, where ${\cal D}$ is the dimension of the object.
The dimension ${\cal D}$ can thus be defined as

\begin{equation}
  {\cal D}=\frac{\log M^{\cal D}}{\log M^{~\ }}=\frac{\log({\rm number\ of\ \mbox{self-similar}\ objects})}{\log({\rm magnification\ factor})} .
\label{dimdef1}
\end{equation}

According to this formula, the dimension of the Sierpinski gasket is 
fractional. When the magnification factor $M$ is 2 there are 3 identical 
pieces of the gasket, for $M=4$ the gasket consists of 9 small copies of 
itself. Therefore its fractal dimension is defined by 

\begin{equation}
{\cal D}=\frac{\log 3 }{\log 2 }=\frac{\log 9 }{\log 4}=1.58496\ldots
\end{equation}

Roughly speaking, the fractal dimension describes how complicated or how large
a self-similar object is. 
A plane is `larger' than a line. The Sierpinski gasket is not a line but also
far from being a plane. Actually, there exist fractals which are constructed
from lines but have dimensions 2 or 3, and therefore fill a plane or a space.
An example of such a curve is the so called Hilbert curve.

The definition of a dimension, given in equation (\ref{dimdef1}), may be 
generalised for the case of non-discrete fractals. In this generalisation, 
the magnification (scaling) factor is a real number $z$ and the number of 
self-similar objects is represented by a density function $f(z)$. Taking 
into account that the dimension may change with scaling, a 
{\it local dimension} is defined as 

\begin{equation}
  {\cal D}(z)=\frac{\partial\log f(z) }{\partial\log z } .
\label{dimdef2}
\end{equation}

For ideal mathematical fractals, discussed so far, ${\cal D}(z)$ is constant 
for the whole fractal. Introducing a scale dependent dimension is natural 
because many fractals in nature (e.g. plants or coastlines) are not 
mathematically ideal and usually have a fractal structure only for a certain 
region of magnification.
In such a region, the dimension is approximately constant, 
${\cal D}(z)={\cal D}$, and, following eq. (\ref{dimdef2}), the density 
function $f(z)$ is given as

\begin{equation}
  \log f(z)={\cal D} \cdot \log z + {\cal D}_0
\label{powerlaw}
\end{equation}
where ${\cal D}_0$ defines the normalisation of $f(z)$, which thus has a {\it power law} behaviour, \mbox{$f(z) \propto z^D$.}

In general, fractals may have two {\it independent} magnification factors, 
$z$ and $y$. In this case the density $f(z,y)$ is written in the following way

\begin{equation}
  \log f(z,y)={\cal D}_{zy} \cdot \log z \cdot \log y + {\cal D}_z \cdot \log z + {\cal D}_y \cdot \log y + {\cal D}_0 .
\label{twod}
\end{equation}

Here the dimension ${\cal D}_{zy}$ represents the dimensional correlation 
relating the $z$ and $y$ factors. The function $f(z,y)$ satisfies a power 
law behaviour in $z$ for fixed $y$ and in $y$ for fixed $z$.

It is important to mention that there is certain freedom in selecting 
magnification factors without changing a shape of the function $f(z,y)$. 
It is possible to use any non-zero power of a factor multiplied by a constant:
$z \rightarrow a z^\lambda$ . The only effect of such a change is a 
redefinition of the dimensional parameters ${\cal D}_{\{z,y,zy\}}$ and of 
the normalisation ${\cal D}_0$, respectively.


\section{Self-similar Structure of the Proton} 
\label{confinement}

\begin{figure}[h]
\centering\epsfig{file=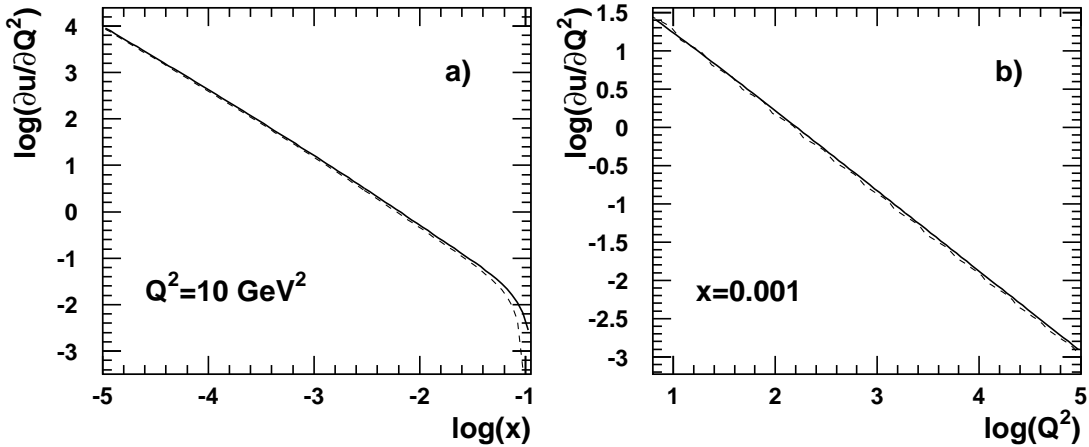,width=15.0cm,bbllx=10pt,bblly=16pt,
bburx=535pt,bbury=235pt}
\caption{Logarithm of the unintegrated u-quark density $\partial u(x,Q^2) / \partial Q^2$ as a function of Bjorken $x$ (a) and $Q^2$ (b). The full and dashed lines correspond to GRV parametrisations in LO and NLO \cite{pdflib}, respectively.} \label{udens}
\end{figure}

Following the dimensional description of the presented fractal structures, 
it is interesting to study the properties of functions describing proton 
structure. In quantum chromodynamics the behaviour of the sea quark densities 
is driven by gluon emissions and splittings. The deeper the proton structure 
is probed, the more gluon-gluon interactions can be observed. These, in 
analogy to fractals, may follow self-similarity, i.e. scaling described by a 
power law. Indeed, there is a number of hints for a self-similar structure. 
As an example, Figure \ref{udens} shows the unintegrated u-quark density for 
fixed momentum transfer $Q^2$ and fixed Bjorken $x$, respectively.

For $x\lesssim 0.01$ (below the valence quark region) the unintegrated 
density function in log-log scale is linear. A linear behaviour is also 
exhibitted by the unintegrated density as a function of $Q^2$ for fixed $x$. 
Refering to eq. (\ref{powerlaw}), this suggests that $x$ and $Q^2$ could be 
treated as appropriate magnification (scaling) factors. This is supporting 
the idea that the proton structure exhibits self-similar properties and may 
be described as a fractal object.

Magnification factors are supposed to fulfil some criteria. They should be 
positive, non-zero and have no physical dimension. The two latter requirements
 concern the selection of $Q^2$ as a magnification factor. The physical 
dimensionality may be removed by dividing $Q^2$ by a constant $Q_0^2$. For 
the case of $Q^2=0$, the non-zero requirement is not fulfilled, however, the 
access to this region is needed for integration of unintegrated densities. 
Thus instead of $Q^2$ a choice of $1+Q^2/Q_0^2$ as a magnification factor is 
appropriate. According to the freedom in the magnification factor selection, 
mentioned above, other equivalent choices are also possible, e.g. 
$\ Q_0^2/(Q_0^2+Q^2)$, $(Q_1^2+Q^2)/1 {\rm GeV}^2$ or similar combinations. 
It is also more appropriate to use $1/x$ as a magnification factor rather 
than $x$ itself: when the structure is probed deeper, $x$ goes to zero 
while a magnification factor should rise.


\section{Structure Function Parametrisation}
\label{parametris}

The concept of self-similarity, when applied to proton confinement structure, 
leads to a simple parametrisation of quark densities within the proton in a 
straightforward way based on Eq. (\ref{twod}). Using magnification factors 
$1/x$ and $1+Q^2/Q_0^2$, an unintegrated quark density may be written in the 
following general form

\begin{equation}
  \log f_i(x,Q^2)={\cal D}_{1} \cdot \log \frac{1}{x} \cdot \log (1+\frac{Q^2}{Q_0^2}) + {\cal D}_2 \cdot \log \frac{1}{x} + {\cal D}_3 \cdot \log (1+\frac{Q^2}{Q_0^2}) + {\cal D}^i_0
\label{param1}
\end{equation}

where $i$ denotes a quark flavour. Conventional, integrated quark densities 
$q_i(x,Q^2)$ are defined as a sum over all contributions with quark 
virtualities smaller than that of the photon probe, $Q^2$.  Thus $f_i(x,Q^2)$ 
has to be integrated over $Q^2$,

\begin{equation}
  q_i(x,Q^2) = \int^{Q^2}_0 f_i(x,q^2)\ dq^2 .
\label{param2}
\end{equation}

Solving equation (\ref{param2}), the following analytical parametrisation of a
quark density is obtained 
\begin{equation}
  q_i(x,Q^2) = \frac{{\mathbf e}^{{\cal D}^i_0}~Q_0^2~x^{-{\cal D}_2}}{1+{\cal D}_3-{\cal D}_1\log {x}}\left(x^{-{\cal D}_1\log (1+\frac{Q^2}{Q_0^2})}(1+\frac{Q^2}{Q_0^2})^{{\cal D}_3+1}-1\right) .
\label{param3}
\end{equation}

Notice that in this parametrisation only the normalisation parameter 
${\cal D}_0^i$ depends on the quark flavour while the other parameters are 
flavour independent. This assumption means that all quarks are following the 
fractal structure, i.e. the dimensions $D_i$ and the magnification factors 
are common for all of them and they differ in normalisation only.

The proton structure function \F is related directly to the quark densities 
$F_2=x \sum_{i}e_i^2(q_i+\bar q_i)$. Thus the assumption about the flavour 
symmetry of Eq. (\ref{param3}) allows to express \F directly in the form 
given on the r.h.s. of Eq. (\ref{param3}) with $x^{-{\cal D}_2}$ replaced by 
$x^{-{\cal D}_2+1}$ and with a common normalisation factor 
${\mathbf e}^{{\cal D}_0}$:

\begin{equation}
  F_2(x,Q^2) = \frac{{\mathbf e}^{{\cal D}_0}~Q_0^2~x^{-{\cal D}_2+1}}{1+{\cal D}_3-{\cal D}_1\log {x}}\left(x^{-{\cal D}_1\log (1+\frac{Q^2}{Q_0^2})}(1+\frac{Q^2}{Q_0^2})^{{\cal D}_3+1}-1\right) .
\label{paramf2}
\end{equation}


\section{Fit to the Data}
\label{fittodata}

\begin{figure}
\centering\epsfig{file=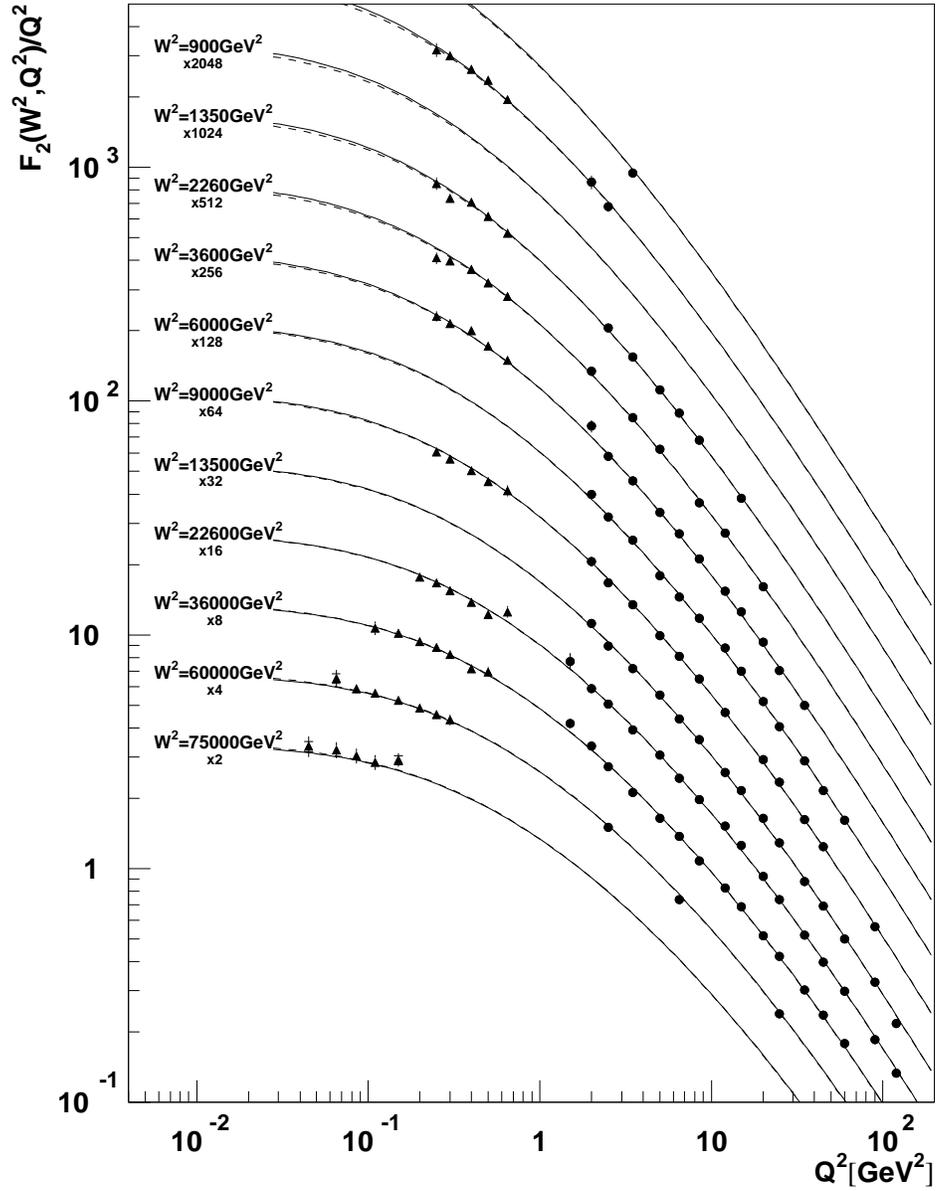,width=13.0cm,bbllx=0pt,bblly=20pt,
bburx=380pt,bbury=490pt}
\caption{Virtual photon-proton cross-section $\sigma_{\gamma^\star p}\propto F_2(W^2,Q^2)/Q^2$ as a function of $Q^2$ in $W^2$ bins. H1 (points) and ZEUS (triangles) measurements are shown along with the fit to 4 parameters (full line) and to all 5 parameters (dashed line).} \label{fith1zeus}
\end{figure}

\begin{figure}[t]
\centering\epsfig{file=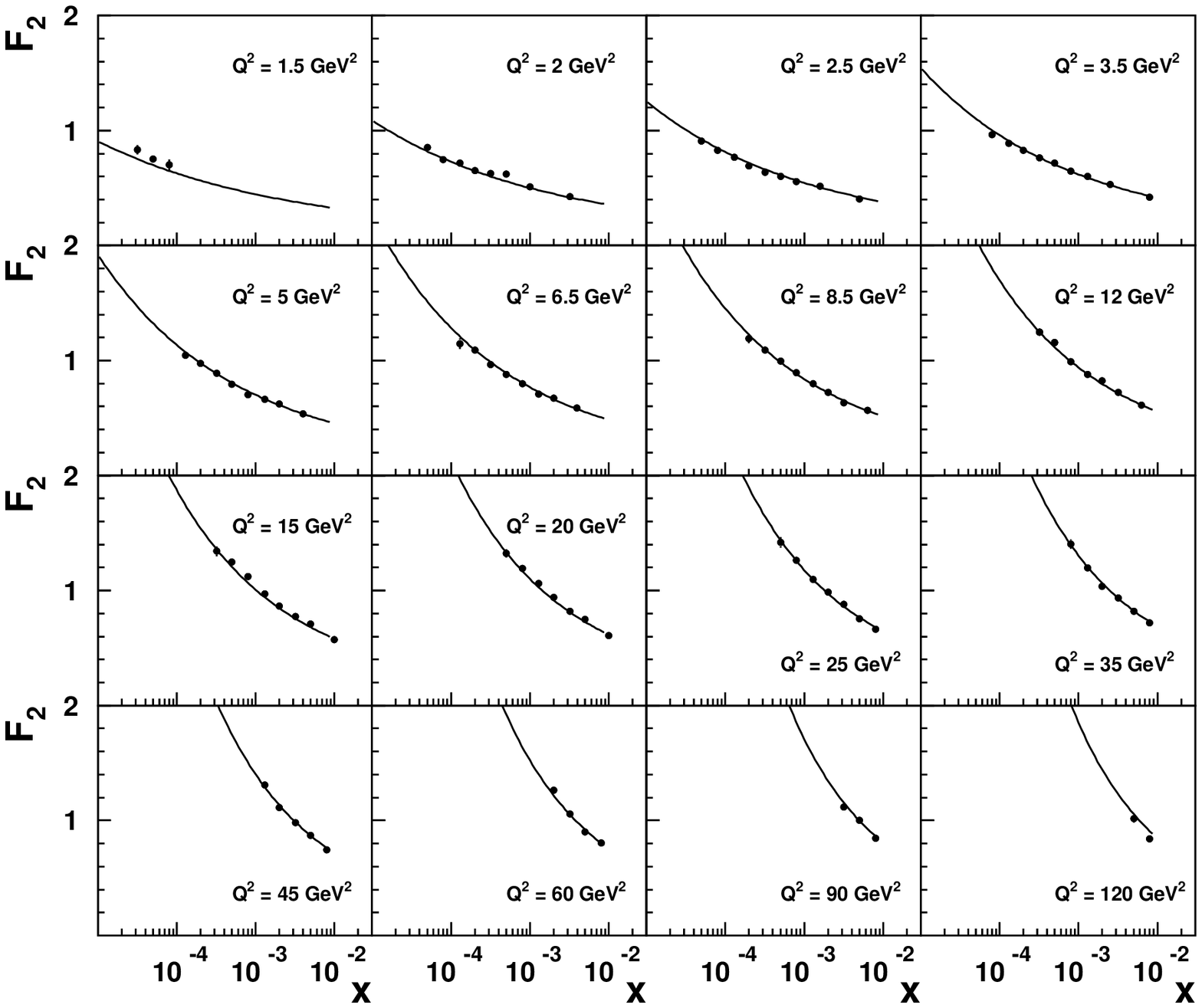,width=14.0cm,bbllx=2pt,bblly=10pt, bburx=524pt,bbury=450pt}
\caption{Measurement of the structure function \F as a function of $x$ in bins of $Q^2$ by the H1 experiment. The curve represents the fit to 4 parameters, which is indistinguishable from the 5 parameter fit in this kinematic region.} \label{fith1}
\end{figure}

\begin{figure}[b]
\centering\epsfig{file=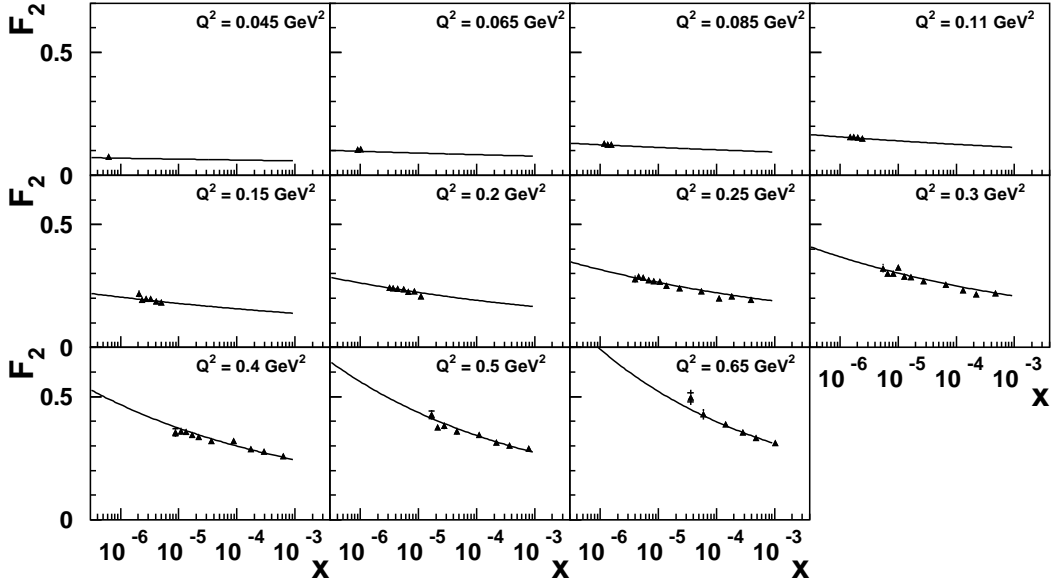,width=14.0cm,bblly=10pt,bbury=310pt,bbllx=2pt,bburx=524pt}
\caption{Measurement of the structure function \F as a function of $x$ in bins of $Q^2$ by the ZEUS experiment. The curve represents the fit to 4 parameters, which is indistinguishable from the 5 parameter fit in this kinematic region.} \label{fitzeus}
\end{figure}

The five parameters ${\cal D}_i$ and $Q_0^2$ are determined using recent 
data from the HERA experiments H1 \cite{thepaper} and ZEUS \cite{f2zeus} in 
the range $1.5 \le Q^2 \le 120 ~{\rm GeV}^2$ (H1) and 
$0.045 \le Q^2 \le 0.65 ~{\rm GeV}^2$ (ZEUS). Additionally a cut $x<0.01$ has
been applied to exclude the valence quark region. The fit parameters are given
in Table \ref{parameters} and the corresponding description
of the \F data is shown in Figures \ref{fith1zeus}, \ref{fith1} and 
\ref{fitzeus}. The $\chi^2$ was calculated with total errors, adding the 
statistical and systematical errors in quadrature. 
When the relative normalisation of the H1 and ZEUS data, which cover
different $Q^2$ regions, was fitted no change beyond 1\% was imposed
by the fits. Thus the normalisations of the data sets were left untouched.

\begin{table}[h]
\centering
\begin{tabular}{|l||c|c|c|c|c||c|c|}
\hline
& ${\cal D}_0$ & ${\cal D}_1$ & ${\cal D}_2$ & ${\cal D}_3$ & $Q_0^2 [{\rm GeV}^2]$ & $\chi^2$ & $\chi^2/ {\rm ndf}$ \\ 
\hline \hline
all fit & 0.339 & 0.073 & 1.013 & -1.287 & 0.062 & 136.6 & 0.82 \\ 
{\scriptsize } & {\scriptsize $\pm 0.145$} & {\scriptsize $\pm 0.001$} & {\scriptsize $\pm 0.01$} & {\scriptsize $\pm 0.01$} & {\scriptsize $\pm 0.01$} & {\scriptsize } & {\scriptsize } \\
\hline
${\cal D}_{2}$ fixed & 0.523 & 0.074 & {\it 1} & -1.282 & 0.051 & 138.4 & 0.82\\
{\scriptsize } & {\scriptsize $\pm 0.014$} & {\scriptsize $\pm 0.001$} & {\scriptsize \it const.} & {\scriptsize $\pm 0.01$} & {\scriptsize $\pm 0.002$} & {\scriptsize } & {\scriptsize } \\
\hline
\end{tabular}
\caption{Results of the fit. The first row corresponds to a fit to all parameters, in the second row parameter ${\cal D}_2$ was fixed to $1$. The number of $F_2$ data points is 172, total errors were used for the $\chi^2$ calculation.}
\label{parameters}

\end{table}

Refering to Figure \ref{fith1zeus} the ratio $F_2(W^2,Q^2)/Q^2$ is proportional
to the virtual photon-proton cross-section $\sigma_{\gamma^\star p}(W^2,Q^2)$.
In the limit $Q^2 \rightarrow 0$ and fixed $W^2$ the parametrisation 
(\ref{paramf2}) behaves like $Q^2$ only for ${\cal D}_2=1$. This may be easily
shown e.g. when the unintegrated structure function $f(x,q^2)$ is introduced

\begin{equation}
  F_2(x,Q^2) = \int^{Q^2}_0 f(x,q^2)\ dq^2
\label{unintf2}
\end{equation}
the parametrisation of which is identical to (\ref{param1}), with ${\cal D}_2$
replaced by ${\cal D}_2-1$ and ${\cal D}^i_0$ replaced by ${\cal D}_0$. If \F 
behaves like $Q^2$ for \mbox{$Q^2 \rightarrow 0$} then $f(x,q^2)$ has to 
behave like a constant for any \mbox{$x = Q^2/(W^2-M_p^2) \rightarrow 0$}. 
That is possible only if the divergent term, involving ${\cal D}_2$, is zero, 
i.e. for \mbox{${\cal D}_2=1$}. In this case, since other logarithmic terms go
to zero, the ratio $F_2(W^2,Q^2)/Q^2$ for $Q^2 \rightarrow 0$ approaches the 
value ${\mathbf e}^{{\cal D}_0}$.

In the fit with ${\cal D}_2$ as a free parameter a value very close to $1$ is 
obtained.  Thus a second fit was made, where ${\cal D}_2$ is fixed to $1$ (see
 Table \ref{parameters}, second row). This fit has 4 
parameters and gives nearly the same $\chi^2/ {\rm ndf}$ as the first fit to 
all 5 parameters. Within the kinematic range of the $F_2$ data, both fits are 
nearly indistinguishable. As was stated above, the parameter ${\cal D}_0$ 
determines the virtual photon-proton cross-section in the photoproduction 
limit. Its value, obtained from the fit, gives 

\begin{equation}
\sigma_{\gamma p}= \left[ \frac{4 \pi^2 \alpha}{Q^2}F_2(W^2,Q^2) \right]_{Q^2 \rightarrow 0} \doteq 189 \pm 3 ~{\rm \mu b}.
\label{gammap}
\end{equation}
This is in approximate agreement with the total photoproduction cross-sections measured by the H1 \cite{gammapH1} and ZEUS \cite{gammapZEUS} collaborations which were not used in the fit. 


\section{Summary}   
\label{summary}

The concept of the self-similar structure of the proton was introduced. This 
leads to a parametrisation of the proton structure function \F which describes
very well the low $x$ HERA data, both in the non-perturbative and the deep 
inelastic domain. The introduced formalism uniquely defines the $x$ and $Q^2$
dependence of parton densities, thus this approach is applicable also to other
measures of proton structure, like the longitudinal structure function $F_L$, 
the diffractive structure function $F_2^D$ or the spin structure function 
$g_1$.                                    

\vspace{0.5cm}
{\bf Acknowledgements}                                                         

I would like to express my gratitude to Max Klein and Krzysztof Golec-Biernat 
for fruitful discussions and careful reading of the manuscript. I am also 
very grateful to my family for the support.


\normalsize   
\noindent 


\end{document}